\documentclass[prd,aps,preprint,tightenlines,superscriptaddress]{revtex4-1}
\usepackage{epsfig}
\usepackage{amsmath}
\begin{document}

\title{QCD Resummation for Single Spin Asymmetries}

\author{Zhong-Bo Kang}
\affiliation{RIKEN/BNL Research Center, Building
510A, Brookhaven National Laboratory, Upton, NY 11973}
\author{Bo-Wen Xiao}
\affiliation{Department of Physics, Pennsylvania State University, University Park, PA
16802, USA}
\author{Feng Yuan}
\affiliation{Nuclear Science Division,
sLawrence Berkeley National Laboratory, Berkeley, CA
94720}
\affiliation{RIKEN/BNL Research Center, Building
510A, Brookhaven National Laboratory, Upton, NY 11973}

\date{\today}
\vspace{0.5in}
\begin{abstract}
We study the transverse momentum dependent factorization for single spin asymmetries in 
Drell-Yan and semi-inclusive deep inelastic scattering processes at one-loop order. 
The next-to-leading order hard factors are calculated in the 
Ji-Ma-Yuan factorization scheme. We further derive the QCD resummation 
formalisms for these observables following the Collins-Soper-Sterman method. 
The results are expressed in terms of the collinear correlation functions from initial and/or final state
hadrons coupled with the Sudakov form factor containing all order soft-gluon resummation effects.
The scheme independent coefficients are calculated up to one-loop order. 
\end{abstract}

\maketitle


Single transverse spin asymmetry in hadronic reactions have 
attracted great attentions from both experiment and theory sides
in recent years. It promises strong connection to the three-dimensional 
partonic tomography of the nucleon, and provides unique 
opportunities to study QCD dynamics, such as the factorization 
and universality of parton distributions. In particular, the short
distance partonic interactions resulting into different Wilson lines
in the relevant transverse momentum dependent (TMD)
parton distributions between the Drell-Yan lepton pair production in $pp$ collisions
and the semi-inclusive hadron production in deep inelastic
scattering (SIDIS) predicts opposite signs for
the Sivers-type single transverse spin asymmetries in these
two processes~\cite{Brodsky:2002cx,Collins:2002kn}. This nontrivial universality prediction
has stimulated strong interests  
around the world to measure, especially, the SSA in the
Drell-Yan (DY) process, since that in the SIDIS has been 
observed in various experiments~\cite{BNLWorkshop}. 

However, the numerical predictions for the SSAs in these processes
are all based on a leading order naive TMD factorization~\cite{Efremov:2004tp}. 
Although a general TMD factorization has been argued for
these processes~\cite{Collins:1981uk,Collins:1984kg,Ji:2004wu,Collins:2004nx,Aybat:2011zv},
we need to know the next-to-leading-order (NLO) perburbative 
QCD corrections to have more reliable predictions, which have not
yet been calculated. Another important issue is the energy
dependence. Current DIS experiments cover the $Q^2$ range 
about 2-5GeV$^2$, whereas that in the planned DY process 
will be measured at relative larger $Q^2$ about 20-25GeV$^2$. 
Here, $Q$ represents the large
momentum scale, i.e., the virtuality of the photon in these
processes. To accurately describe the $Q^2$ evolution of the transverse
momentum dependent observables, the QCD resummation effects
have to be taken into account~\cite{Collins:1984kg}.

In this paper, we will build a theoretical framework to address these
important questions. We carry out, at the first time,
the complete NLO perturbative correction to 
the single transverse spin dependent cross sections in 
DY and SIDIS processes in the TMD factorization. 
One of the important implications of the explicit one-loop
calculations is to help to construct the {\it correct} resummation
formalism for the SSA observables.
Earlier attempts to formulate these effects in 
SSAs have been made in various 
forms~\cite{Boer:2001he,Idilbi:2004vb}.
A resummation formula close to ours was used in 
Ref.~\cite{Boer:2001he}, and a significant suppression effect
were found when $Q^2$ is very large. In the following,
based on the one-loop calculation results, 
we will derive the complete soft gluon resummation 
formalism  following the Collins-Soper-Sterman (CSS) 
method~\cite{Collins:1984kg}, which can be 
easily implemented in the phenomenological 
studies.

We take the DY process as an example to demonstrate our
procedure and present the main results,
\begin{equation}
A (P_A,S_\perp)+B(P_B) \to \gamma^* (q) +X \to \ell^+ + \ell^ -
+X,
\end{equation}
where $P_A$ and $P_B$ represent the momenta of hadrons $A$ and $B$, 
and $S_\perp$ for the transverse polarization vector of $A$,
respectively. The single transverse spin dependent 
differential cross section can be expressed as
\begin{eqnarray}
\frac{d\Delta\sigma(S_\perp)}{dydQ^2d^2q_\perp}&=&
\sigma_0\epsilon^{\alpha\beta}S_\perp^\alpha W_{UT}^\beta (Q;q_\perp)\ ,
\end{eqnarray}
where $q_\perp$ and $y$ are transverse momentum and rapidity of the lepton pair, 
$\sigma_0=4\pi\alpha_{em}^2/3N_csQ^2$ with $s=(P_A+P_B)^2$,
and $\epsilon^{\alpha\beta}$ is defined 
as $\epsilon^{\alpha\beta\mu\nu}P_{A\mu}P_{B\nu}/P_A\cdot P_B$.
At low transverse momentum ($q_\perp\ll Q$) the structure function $W_{UT}$ can be formulated 
in terms of the TMD factorization where the quark Sivers function is
involved~\cite{Brodsky:2002cx,Collins:2002kn}, 
whereas at large transverse momentum ($q_\perp\gg \Lambda_{\rm QCD}$ )
it can be calculated in the collinear factorization approach in terms of the twist-three
quark-gluon-quark correlation functions~\cite{Ji:2006ub,Efremov:1981sh,Qiu:pp}.
It has been shown that the TMD and collinear twist-three approaches
give the consistent results in the intermediate transverse momentum region:
$\Lambda_{\rm QCD}\ll q_\perp\ll Q$~\cite{Ji:2006ub,bbdm}. This consistency allows 
us to separate $W_{UT}$ into two terms~\cite{Collins:1984kg},
\begin{equation}
W_{UT}^\alpha(Q;q_\perp)=\int \frac{d^2b}{(2\pi)^{2}} e^{i \vec{q}_\perp\cdot \vec{b}} \widetilde
{W}_{UT}^\alpha(Q;b) +Y_{UT}^\alpha(Q;q_\perp) \ , \nonumber
\end{equation}
where the first term dominates at $q_\perp\ll Q$ region, and the
second term at $q_\perp\sim Q$. The latter is obtained
by subtracting the low $q_\perp$ expansion from 
the full perturbative calculation. 

According to the TMD factorization, we have~\cite{Ji:2004wu,Idilbi:2004vb}, 
\begin{eqnarray}
\widetilde{W}_{UT}^\alpha(Q;b)&=&\tilde f_{1T}^{(\perp\alpha)}(z_1,b,\zeta_1)
\bar q(z_2,b,\zeta_2)\nonumber\\
&&\times H_{UT}(Q)\left(S(b,\rho)\right)^{-1} \ ,
\end{eqnarray}
where $z_{1,2}=Q/\sqrt{s}e^{\pm y}$ and the sums over flavor weighting 
with the charge squared of the quarks are implicit, $f_{1T}^\perp$ and $\bar q$
are the TMD quark Sivers function of $A$ and antiquark distribution of $B$, 
$H$ and $S$ are hard and soft factors, respectively. In this paper, 
we follow the Ji-Ma-Yuan factorization scheme~\cite{Ji:2004wu}, where 
two off-light-front Wilson lines (along off-light-front vectors $v_1$ and $v_2$)
are introduced to regulate the light-cone singularities for the TMD quark distributions. 
We further define $\zeta_1^2=(v_1\cdot P_A)^2/v_1^2$ and 
$\zeta_2^2=(v_2\cdot P_B)^2/v_2^2$, and the rapidity cut-off parameter $\rho$:
$z_1^2\zeta_1^2=z_2^2\zeta_2^2=\rho Q^2$. The Sivers function in 
the impact parameter $b_\perp$-space is defined
as $\tilde f_{1T}^{(\perp\alpha)}(x,b_\perp)=\int d^2k_\perp e^{-i\vec{k}_\perp\cdot \vec{b}_\perp}
k_{\perp}^\alpha f_{1T}^{\perp({\rm DY})}(x,k_\perp)/M_P$,
where $f_{1T}^\perp$ follows the definition of Ref.~\cite{Bacchetta:2006tn}. 
Our results can be translated to other factorization schemes where 
different regularizations for the light-cone singularities are used~\cite{Aybat:2011zv}.

We investigate the above factorization formula Eq.~(3) in 
the perturbative region of $1/b_\perp\gg \Lambda_{\rm QCD}$. 
The explicit calculations at one-loop order
will verify the TMD factorization,  
from which we obtain the NLO correction to the hard factor. 
In the calculation of the spin-average (or double spin asymmetry)
TMD factorization, it is convenient to choose a quark (or gluon) 
target, where every factor in the factorization formula can be 
calculated perturbatively~\cite{Ji:2004wu}. 
However, the SSA vanishes with on-shell quark. 
To get nonzero effect, we have to go beyond the simple quark
target picture. 

Our calculations are based on the collinear correlation functions
from the incoming hadrons. In this framework, the SSA is naturally 
a twist-three effect, and involves the twist-three quark-gluon-quark 
correlation function from the polarized nucleon~\cite{Efremov:1981sh,Qiu:pp}. 
There have been great theoretical developments 
in this framework in recent years, see, e.g., Refs.~\cite{Kouvaris:2006zy,
Ji:2006ub,Eguchi:2006mc,Vogelsang:2009pj,Kang:2008ey}. We will utilize these
techniques to compute the structure function $\widetilde{W}_{UT}^\alpha(Q,b)$.
First, let us write down a general form,
\begin{eqnarray}
\widetilde{W}_{UT}^\alpha(Q,b)&=&\left(\frac{-ib_\perp^\alpha}{2}\right)
\int\frac{dx_1dx_2dx'}{x_1x_2x'}T_{F}(x_1,x_2)\bar q(x')\nonumber\\
&&\times{\cal H}(x_1,x_2,x';Q,b) \ ,
\end{eqnarray}
where  $T_{F}$ is the Qiu-Sterman matrix element for 
quark $q$~\cite{Qiu:pp,Vogelsang:2009pj}
and $\bar q(x')$ the integrated anti-quark distribution.
The quark Sivers function is related to the Qiu-Sterman
matrix element: $T_F(x,x)=\int d^2k_\perp|k_\perp^2|
f_{1T}^{\perp({\rm DY})}(x,k_\perp)/M_P$~\cite{Ji:2006ub,Vogelsang:2009pj}. 
Other twist-three quark-gluon-quark correlation functions contributing
to the SSA can be included as well. For simplicity,
we focus on $T_F$ contributions in this paper.

\begin{figure}[t]
\begin{center}
\includegraphics[width=7cm]{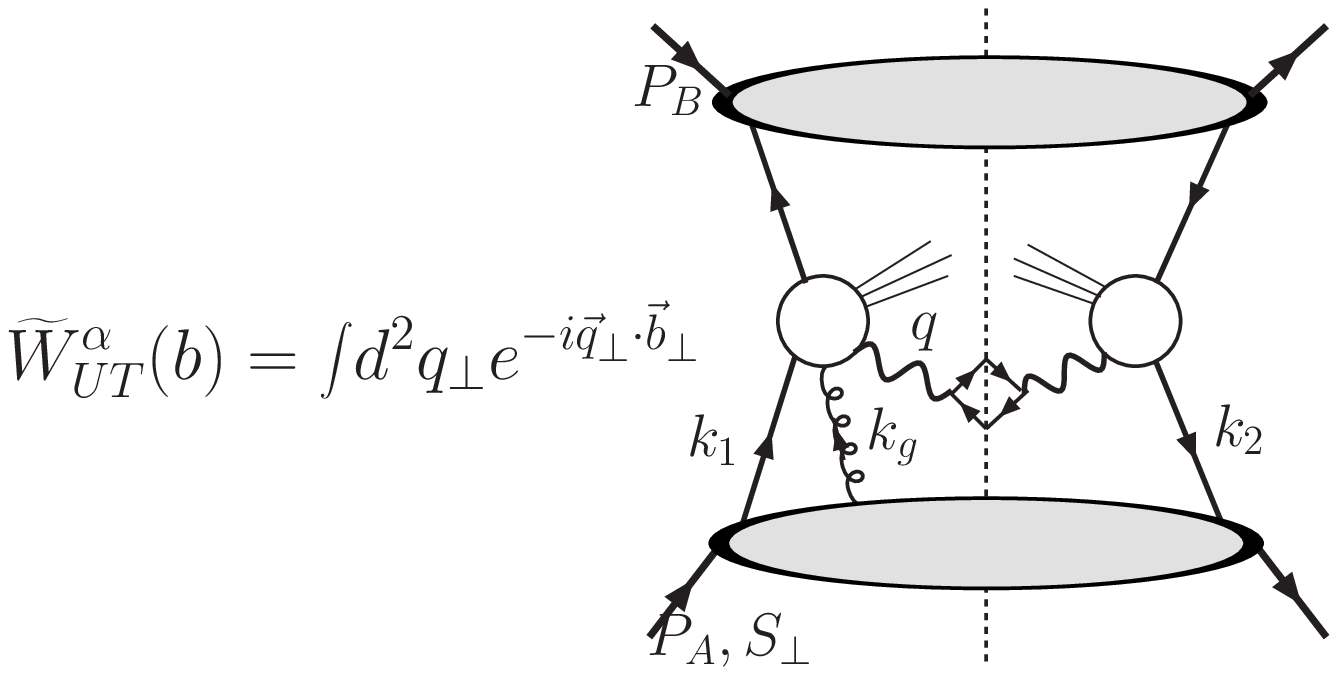}
\end{center}
\vskip -0.4cm \caption{\it Generic Feynman diagram contribution
to the impact parameter space structure function $\widetilde{W}_{UT}(Q,b)$.}
\end{figure}

The above formula is similar to that of the $q_\perp$-weighted asymmetry in 
the same process calculated in Ref.~\cite{Vogelsang:2009pj}. 
There are two major differences: (1) instead of weighting with $q_\perp^\alpha$, here
we weight with $e^{-i\vec{q}_\perp\cdot \vec{b}_\perp}$; (2) we will restrict our calculations
of the differential cross section in the TMD domain, i.e., $q_\perp\ll Q$ 
($1/b_\perp\ll Q$). We calculate Eq.(4) in a covariant gauge, where a generic diagram
is illustrated in Fig.~1. 
Because of the twist-three effect, an additional gluon
attachment has to be taken into account for hard partonic 
scattering amplitude. 
A collinear expansion will be performed to calculate the hard
part ${\cal H}(Q,b)$. In particular, the 
amplitude is expanded in terms
of $k_{g\perp}^\alpha=k_{2\perp}^\alpha-k_{1\perp}^\alpha$
where $k_{i\perp}\ll 1/b_\perp$. 
Combining this expansion with matrix element from the polarized
nucleon, it will lead to the
spin dependent cross section expressed in terms of $T_F(x_1,x_2)$.

The leading order diagrams are shown in Fig.~2(a) and (b).
From the kinematics,  we find that $q_\perp$ is related to the 
transverse momenta of the two quark lines as:
$q_\perp=k_{2\perp}$ for Fig.~2(a) and $q_\perp=k_{1\perp}$
for Fig.~2(b). Therefore, the contributions from these two
diagrams will be
\begin{eqnarray}
{\rm Fig.2}&=&\int d^2q_\perp e^{-i\vec{q}_\perp\cdot \vec{b}_\perp}\left(
\frac{ig}{-(k_{2}^+-k_{1}^+)-i\epsilon}\right)\nonumber\\
&&\times\left[\delta(q_\perp-k_{2\perp})-\delta(q_\perp-k_{1\perp})\right] \nonumber\\
&=& \left(\frac{ig}{-(k_{2}^+-k_{1}^+)-i\epsilon}\right)
\left[e^{-i\vec{k}_{2\perp}\cdot \vec{b}_\perp}-e^{-i\vec{k}_{1\perp}\cdot \vec{b}_\perp}\right]
\nonumber\\
&=&\frac{ig}{-(k_{2}^+-k_{1}^+)-i\epsilon}\left(-ib_\perp^\alpha\right) k_{g\perp}^\alpha \ ,
\label{eq7}
\end{eqnarray}
where the last equation comes from the collinear expansion of 
the exponential factor. The initial state interactions
represented by the propagator leads to 
a pole contribution. After taking the pole, we will
obtain the leading order contribution,
\begin{equation}
\widetilde{W}_{UT}^{\alpha(0)}(Q,b)=\left(\frac{-ib_\perp^\alpha}{2}\right)
T_F(z_1,z_1)\bar q(z_2) \ .
\end{equation}
This also normalizes the leading order hard factor as $H_{UT}^{(0)}=1$
in Eq.~(3), because $f_{1T}^{\perp\alpha({\rm DY})}(z_1,b_\perp)=T_F(z_1,z_1)(-ib_\perp^\alpha/2)$
at this order at small $b_\perp$.

\begin{figure}[t]
\begin{center}
\includegraphics[width=9cm]{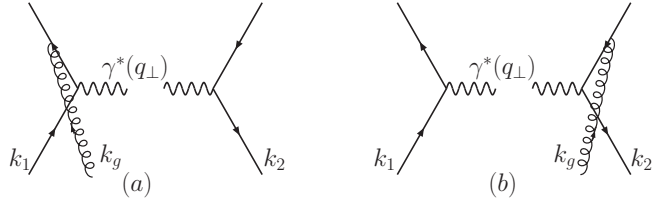}
\end{center}
\vskip -0.4cm \caption{\it Leading order Born diagram 
calculation of $\widetilde{W}_{UT}(Q,b)$.}
\end{figure}

Order $\alpha_s$ corrections come from real and virtual
gluon radiation contributions. Similar to that calculated in 
Ref.~\cite{Vogelsang:2009pj}, the virtual diagrams contain
soft and collinear divergences.
The soft divergence will be cancelled out by the
real gluon radiation contributions, and the collinear
divergences will be absorbed into the parton distributions
in both TMD and collinear factorizations in Eqs.~(3) and (4).
The real diagrams are similar to those calculated in 
Refs.~\cite{Ji:2006ub,Vogelsang:2009pj}. In order to obtain
their contributions to $\widetilde{W}_{UT}^\alpha(b_\perp)$, we need
to first take the low transverse momentum limit, and 
then Fourier transform into the impact parameter
$b_\perp$-space. 
After summing up both real and virtual diagrams contributions,
we indeed find that the soft divergences cancel out between each 
other, and the total contribution only contains the collinear
divergences,
\begin{eqnarray}
&&\frac{\alpha_s}{2\pi}\left(\frac{-ib_\perp^\alpha}{2}\right)\left\{
\left[-\frac{1}{\epsilon}+\ln\frac{4}{b^2\mu^2}e^{-2\gamma_E}\right]
\left({\cal P}_{q/q}\otimes \bar q(z_2')\right.\right.\nonumber\\
&&\left.~~+{\cal P}^T_{qg\to qg}\otimes
T_F(z_1')\right)
+C_F(1-\xi_2)\delta(1-\xi_1)\nonumber\\
&&~~+\left(-\frac{1}{2N_c}\right)(1-\xi_1)\delta(1-\xi_2)
+\delta(1-\xi_1)\delta(1-\xi_2)\nonumber\\
&&~~\left.\times\left[-\ln^2\left(\frac{Q^2b^2}{4}
e^{2\gamma_E-\frac{3}{2}}\right)-\frac{23}{4}+\pi^2\right]\right\} \ ,
\end{eqnarray}
where $\xi_i=z_i/z_i'$, ${\cal P}_{qq}$ is the quark splitting function,
and ${\cal P}_{qg\to qg}^T$ the splitting function for the Qiu-Sterman
matrix element~\cite{Vogelsang:2009pj,Kang:2008ey}. After subtracting the collinear divergences from 
the splitting of $T_F$ and $\bar q$, we demonstrate the factorization 
form in Eq.~(4) up to one-loop order.

The above result can also be casted into the TMD factorization
formula Eq.~(3), where we have to subtract the TMD quark Sivers
function, antiquark distribution and the soft factor. The latter two have
been calculated before~\cite{Ji:2004wu}. The quark Sivers function
can be calculated similarly, and we find that,
\begin{eqnarray}
\!\!&&\!\!\tilde{f}_{1T}^\alpha(z_1,b_\perp)=\frac{\alpha_s}{2\pi}\left(\frac{-ib_\perp^\alpha}{2}\right)
\left\{
\left[-\frac{1}{\epsilon}+\ln\frac{4}{b^2\mu^2}e^{-2\gamma_E}\right]
\right.\nonumber\\
&&\times{\cal P}^T_{qg\to qg}\otimes
T_F(z_1')+\delta(1-\xi_1)C_F\left[-\frac{3}{2}\ln\frac{4}{b^2\mu^2}e^{-2\gamma_E}\right.\nonumber\\
&&\left.\left.-\frac{1}{2}\ln^2\left(\frac{z_1^2\zeta_1^2b^2}{4}
e^{2\gamma_E-1}\right)
-\frac{3+\pi^2}{2}\right]\right.\nonumber\\
&&\left.
+\left(-\frac{1}{2N_c}\right)(1-\xi_1)
\right\} \ .
\end{eqnarray}
After subtracting these factors out,  
we find that the hard factor $H_{UT}$ is free of infrared divergence,
and it is  the same as that for the spin average one calculated in Ref.~\cite{Ji:2004wu}, 
\begin{eqnarray}
&&\!\!H_{UT}^{(1){\rm DY}}=H_{UU}^{(1)}|_{\rm DY}
\nonumber\\
&&=\frac{\alpha_s}{2\pi}C_F\left[\ln\frac{Q^2}{\mu^2}(1+\ln\rho^2)
-\ln\rho^2+\ln^2\rho+2\pi^2-4\right],\nonumber
\end{eqnarray}
which is very interesting and suggests that the hard factors 
are spin-independent. Following the same procedure, we find that the hard factor for
the single transverse spin dependent cross section in SIDIS. Again,
it is the same as the spin-average case,
$H_{UT}^{(1)}|_{\rm SIDIS}=H_{UU}^{(1)}|_{\rm SIDIS}$~\cite{Ji:2004wu}.
Clearly, the hard factors for DY and SIDIS differ by a factor of $\pi^2$.
The hard factors in other TMD factorization
scheme can be calculated similarly.

The factorization formula Eq.~(3) contains large logarithms in terms
of $\ln^2 (Q^2b^2)$~\cite{Idilbi:2004vb} which is also shown in the 
total contribution result of Eq.~(7). 
To resum these large logarithms, we follow the CSS method~\cite{Collins:1984kg},
\begin{eqnarray}
\widetilde {W}_{UT}^\alpha(Q;b)&=& e^{-{\cal S}_{UT}(Q^2,b)}\widetilde{W}_{UT}^\alpha(C_1/b,b)
\nonumber\\
&=&\left( {-i b_\perp^\alpha}/{2}\right)e^{-{\cal S}_{UT}(Q^2,b)}\Sigma_{i,j}\nonumber\\
&&\times
\Delta C_{qi}^T\otimes f_{i/A}^{(3)}(z_1) C_{\bar qj}\otimes f_{j/B}(z_2) \ ,
\end{eqnarray}
where $f_{j/B}$ represents the integrated parton distribution from hadron $B$, 
and $f_{i/A}^{(3)}$ the twist-three function from hadron $A$ ($T_F(z_1,z_2)$ 
is the most relevant one). The last step of the above equation comes from further
applying the collinear factorization formula Eq.~(4) at lower energy 
scale $C_1/b$. 
From the above one-loop calculations, we find that 
the perturbative Sudakov factor ${\cal S}_{UT}$ have the same form
as that for the spin-average case,
\begin{eqnarray}
{\cal S}_{UT}(Q^2,b)&=&\int_{C_1^2/b^2}^{C_2^2Q^2}\frac{d\mu^2}{\mu^2}
\left[\ln\left(\frac{C_2^2 Q^2}{\mu^2}\right)A_{UT}(C_1;g(
\mu))\nonumber\right.\\
&&\left.~~+B_{UT}(C_1,C_2;g(\mu))\right] \ ,
\end{eqnarray}
up to one-loop order, where $C_1$ and $C_2$ are constants in the order of $1$. 

The $A$, $B$ and $C$ functions can be calculated in perturbation 
theory: $A=\sum_{n=1}A^{(n)}(\alpha_s/\pi)^n$. 
From the explicit one-loop calculations, we
obtain the following results for these coefficients,
\begin{eqnarray}
A_{UT}^{(1)}&=&C_F,~B_{UT}^{(1)}=-3/2C_F,~\Delta C_{qq}^{T(0)}=\delta(1-x) \ , \nonumber\\
\Delta C_{qq}^{T(1)}&=&-\frac{1}{4N_c}(1-x)+\frac{C_F}{2}\delta(x-1)\left[\frac{\pi^2}{2}
    -4\right] \ ,\nonumber
\end{eqnarray}
where $C_{qq}$ follows the spin-average case~\cite{Collins:1984kg}, and
we have chosen the canonical values for $C_1=2e^{-\gamma_E}$ and $C_2=1$
to simplify the above expressions. For SIDIS, $A$, $B$ remain the same,
whereas $\Delta C^T$ have opposite sign and there is no 
$\pi^2$ term in $\Delta C_{qq}^{T(1)}$.
The above coefficients can also be calculated by 
comparing the fixed order calculations of the differential 
cross section depending on transverse momentum~\cite{Ji:2006ub}
to the expansion of the resummation formula Eq.~(9).
We have checked that this gives the consistent results.

Eq.(9) is our main result for the QCD resummation for the single spin
asymmetry in the DY process. 
The structure, in particular, the pre-factor $(ib_\perp^\alpha)$ is
the unique feature for the single transverse spin azimuthal angel 
dependent cross section. This pre-factor comes from the 
explicit one-loop calculation above. It also guarantees
the proper behaviors for the spin asymmetry 
at small and large transverse momentum. At very small $q_\perp$, the 
asymmetry has to vanish, so that the differential cross section and $W_{UT}^\alpha$
will be proportional to $q_\perp^\alpha$. On the other hand,
at large $q_\perp$, the asymmetry is power suppressed by $1/q_\perp$
from the perturbative calculation, which leads to 
$W_{UT}^\alpha\propto q_\perp^\alpha/q_\perp^4$. With the correct 
coefficients extracted from our one-loop calculation, 
these behaviors are satisfied in Eq.~(9). 
We would like to emphasize that the TMD and collinear 
factorizations for $\widetilde{W}_{UT}^\alpha(B)$
in Eqs.~(3) and (4) are crucial to obtain the final resummation 
formula of Eq.~(9). Without these factorization results, we can not
apply the CSS resummation.

Another important feature is that Eq.~(9) depends on the integrated parton distributions, 
which are universal. The opposite sign for the SSAs in DY and SIDIS
is reflected by the opposite $C$ coefficients in this formula. 
Moreover, our resummation formula is 
scheme-independent, although the TMD factorization Eq.~(3) 
depends on the scheme of how to regulate the 
light-cone singularities in the TMD distributions. 
This can be clearly seen from the disappearance of $\rho$
and $\zeta_i^2$ in Eq.~(9) with he above coefficients. 


In summary, we have derived the CSS resummation
formalism for the single spin asymmetries in DY
and SIDIS processes. The relevant 
coefficients are calculated up to one-loop order. These
results shall be further studied to understand the energy
dependence of the SSAs in these processes, and provide
more accurate predictions for the DY process
which is actively pursued by several experiments.
Our results shall shed light on all other $k_\perp$-odd observables,
and should be applied to the  azimuthal angular dependent 
observables in DY, SIDIS, and $e^+e^-$ annihilation processes. 
We performed our calculations in the framework of
the collinear correlation functions of hadrons. This allows
us to compute the cross sections in a consistent and rigorous way.
We noticed that recently, a different framework has been 
developed where a gluonic degree of freedom is included
in the leading order quark target~\cite{Ma:2011nd}. It will be interesting 
to apply this method and compare with our results.

This work was supported in part by the U.S. Department of Energy
under grant number DE-AC02-05CH11231.  
We are grateful to RIKEN, Brookhaven National Laboratory, 
and the U.S. Department of Energy (Contract No.~DE-AC02-98CH10886) 
for supporting this work.

\end{document}